\documentclass[conference]{IEEEtran}
\IEEEoverridecommandlockouts
\usepackage{cite}
\usepackage{amsmath,amssymb,amsfonts}
\usepackage{algorithmic}
\usepackage{graphicx}
\usepackage{textcomp}
\usepackage{xcolor}
\usepackage{framed}
\def\BibTeX{{\rm B\kern-.05em{\sc i\kern-.025em b}\kern-.08em
    T\kern-.1667em\lower.7ex\hbox{E}\kern-.125emX}}
\begin{document}

\title{Poster: No safety in numbers: traffic analysis of sealed-sender groups in Signal \\
\thanks{Supported by NSF grant 1814753}
}


\author{\IEEEauthorblockN{Eric Brigham}
\IEEEauthorblockA{\textit{Department of Computer Science} \\
\textit{University of Minnesota}\\
Minneapolis, United States \\
brigh169@umn.edu}
\and
\IEEEauthorblockN{Nicholas Hopper}
\IEEEauthorblockA{\textit{Department of Computer Science} \\
\textit{University of Minnesota}\\
Minneapolis, United States \\
hoppernj@umn.edu}
}

\maketitle


\begin{IEEEkeywords}
Signal, end-to-end encrypted messaging applications, statistical disclosure attacks
\end{IEEEkeywords}
\section{Introduction}
Secure messaging applications often offer privacy to users by protecting their messages from would be observers through end-to-end encryption techniques \cite{moxieX3dh} \cite{moxieSesame} \cite{unger2015sok}. 
However, the {\em metadata} of who communicates with whom cannot be concealed by encryption alone.
Signal's Sealed Sender mechanism attempts to enhance its protection of this data by obfuscating the sender of any message sent with the protocol \cite{jlund2018SS}. However, it was shown by Martiny {\em et al.} \cite{martiny2021improving} that due to the message delivery protocols in Signal, the record of who {\em receives} messages can be enough to recover this metadata.  In this work we extend the attack presented in \cite{martiny2021improving} from deanonymizing communicating pairs to deanonymizing entire group conversations.

\section{Background}
Statistical disclosure attacks, or SDA, are a known attack vector which can be used to link senders and receivers of messages in anonymous mix networks \cite{danezis2003statistical} \cite{danezis2007two} \cite{mathewson2004practical} \cite{troncoso2008perfect}. 
The traditional attack methods rely on two key components that seemingly make them ineffective in an environment like Signal: identities of senders are known and there is a mix entity present \cite{chaum1981untraceable}. Under Signal's Sealed Sender mechanism, the identity of a message's sender would be hidden, and there is no mix present. The authors of \cite{martiny2021improving} argue that in the presence of immediate responses, both of these apparent shortcomings can be overcome and a modified SDA attack can be developed and leveled against the users by the server in order to deanonymize communicating pairs. Crucially, they observe that immediate responses are guaranteed in Signal due to the fact that delivered receipts cannot be disabled by the user. The authors show both convincing theoretical analysis as well as simulations that the attack does achieve its goal. In this work we extend their proposed attack from the setting of communicating pairs to that of groups, where there are three or more users communicating through a single channel.

\section{The Attack}
We consider an attack setting where a single user Bob is being targeted in the hope of discovering his group $G$ of $k$ associates who are communicating together through a single channel. 
In Signal, when Bob sends a message through a group channel, within a short period of time which we refer to as the \textit{epoch}, all members of the group can be seen to have received a message, and shortly thereafter, Bob will receive delivered receipts from all members; we refer to this sequence of delivered receipts as the \textit{flurry}.  Contrasting this with a random epoch of the same length where Bob has not sent a group message, the group members may or may not be receiving messages, but Bob will \textit{not} receive a quick succession of messages. We observe then that by monitoring who receives messages most often before a flurry, we can identify the likely members of the group. 

This behavior was confirmed by examining output logs generated by group messaging with the Android emulator Genymotion similarly to \cite{cremers2023formal}. 
While the proposed attack generalizes to a group of any size, we use a group of three members for simplicity of exposition. We assume that all members of the group are online at the time of the attack, otherwise delivered receipts cannot be sent. Similarly to \cite{martiny2021improving} we define random and target epochs, but the foundation of our attack is the \textit{flurry}. The following definition/terms will be used throughout:
\begin{itemize}
    \item \textbf{Flurry} a string of ‘To Bob’ messages which appear to the server as being sent in succession, or very close to it. This occurs after Bob sent a message through the group channel, after which we would necessarily observe delivered receipts to Bob from all members of the group. Please refer to Fig. 1 for a visual representation. 
    \item \textbf{Attack Window} : the time-frame in which the attack takes place, necessarily spanning multiple messages sent to and from Bob
    \item \textbf{Target Epoch}: an epoch during the attack window preceding a flurry of ``To Bob” delivered receipts. This contains some number of messages sent by Bob using sealed sender with all recipients(group members) observable. 
    \item \textbf{Random Epoch}: an epoch taken from the total set of messages, chosen uniformly at random independent from Bob’s activities. 
\end{itemize}

\begin{figure}[htbp]
\centerline{\includegraphics[scale=0.5]{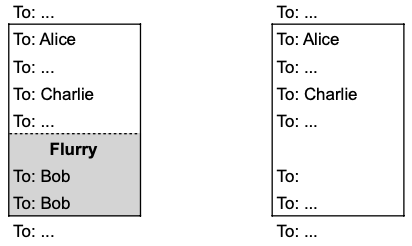}}
\caption{The left box corresponds to a target epoch due to the existence of the flurry, while the right box could correspond to some random epoch}
\label{fig}
\end{figure}
We define our attack on the assumption that a flurry exists. The attack follows the basic functioning presented in \cite{martiny2021improving}, but is modified for a group setting and predicated on the existence of a flurry. It is as follows:  
\begin{enumerate}
    \item Create an empty table of counts, initializing all values to zero. 
    \item Sample a target epoch. For each user that received a message during the target epoch, increment their count in the table. 
    \item Sample a random epoch. For each user that received a message during the random epoch, decrement their count in the table.
    \item Repeat steps two and three for some number $n$ target and random epochs (same amount each). 
    \item The users in the table with the highest counts are most likely to be in a group together.  Note that because a flurry is always produced when Bob sends to the group, but is unlikely to occur in pairwise communication patterns, the process will produce the highest scores for group members communicating with Bob.
\end{enumerate}
Note it possible to define an attack on the absence of a flurry, for instance if Bob were a passive observer of the group, but this is left to a full version of the paper. 

\section{Theoretical Analysis}
The subsequent analysis will rely on the following assumptions, of which one, two, and four are identical to those found in \cite{martiny2021improving}, the third and fifth are unique to this setting:
\begin{enumerate}
    \item The probability of receiving a message during a particular epoch is independent of receiving a message during any other epoch
    \item Each user has a fixed probability of receiving a message during a random epoch, call it $r_u$
    \item Any group member $u$ has a fixed probability $t_u$ of receiving a message during a target epoch, with $t_u > r_u$
    \item All users not in the group have the same probability of receiving a message during a target or random epoch, $t_u = r_u$
    \item Bob is in \textbf{one} group (Note that in this case, assuming {\em no} false positive flurries, we have $t_u = 1$ for all $u \in G$.)
\end{enumerate}

In this section we present theoretical analysis of attack success. It follows much in the same way as the analysis presented in \cite{martiny2021improving}, but for a group setting. \\
\textbf{Theorem:} Given $m$ total users in a messaging system. Let each group member $u \in G$  have probabilities $r_u, t_u$ of appearing in random or target epochs respectively. Then under the stated probability assumptions, the probability that all $k$ group members are ranked higher than all non-associates after $n$ random and target epochs is at least:
$$
1 - \frac{m|G|}{C^n}
$$
Where the parameter $C = \min_{u \in G} exp((t_u - r_u)^2 / 4) > 1$,
depend solely on group member probabilities $r_u,t_u$. The proof of this theorem is left to a full version of the paper. \\
The consequences of this are similar to those found in \cite{martiny2021improving}, chief among them being that the number of epochs needed to de-anonymize group members with high probability scales logarithmically with the total number of users -- meaning that the attack is efficient.  


\section{Future work}
While all that has been included here are theoretical results, simulations are in progress and will be included for a full paper. We also hope to explore the dynamics of the target user being in multiple groups simultaneously. Finally, we hope to develop and test defense mechanisms that do not require Signal to modify their own implementation. 




\bibliographystyle{plain}
\bibliography{reference}


\vspace{12pt}
\color{red}

\end{document}